# Classification of Diabetic Retinopathy Using Unlabeled Data and Knowledge Distillation

*Sajjad Abbasi[1], Mohsen Hajabdollahi[1], Pejman Khadivi[2], Nader Karimi[1], Roshanak Roshandel[2], Shahram Shirani[3], Shadrokh Samavi[1,3]*

[1]Department of Electrical and Computer Engineering, Isfahan University of Technology, Iran
[2]Computer Science Department, Seattle University, Seattle, USA
[3]Department of Electrical and Computer Engineering, McMaster University, Canada

**ABSTRACT**

Over the last decade, advances in Machine Learning and Artificial Intelligence have highlighted their potential as a diagnostic tool in the healthcare domain. Despite the widespread availability of medical images, their usefulness is severely hampered by a lack of access to labeled data. For example, while Convolutional Neural Networks (CNNs) have emerged as an essential analytical tool in image processing, their impact is curtailed by training limitations due to insufficient availability of labeled data. Transfer learning enables models developed for one task to be reused for a second task. Knowledge distillation allows transferring knowledge from a pre-trained model to another. However, it suffers from limitations, and constraints related to the two models need to be architecturally similar. Knowledge distillation addresses some of the shortcomings associated with transfer learning by generalizing a complex model to a lighter model. However, some parts of the knowledge may not be distilled by knowledge distillation sufficiently. In this paper, a novel knowledge distillation approach using transfer learning is proposed. The proposed method transfers the entire knowledge of a model to a new smaller one. To accomplish this, unlabeled data are used in an unsupervised manner to transfer the maximum amount of knowledge to the new slimmer model. The proposed method can be beneficial in medical image analysis, where labeled data are typically scarce. The proposed approach is evaluated in the context of classification of images for diagnosing Diabetic Retinopathy on two publicly available datasets, including Messidor and EyePACS. Simulation results demonstrate that the approach is effective in transferring knowledge from a complex model to a lighter one. Furthermore, experimental results illustrate that the performance of different small models is improved significantly using unlabeled data and knowledge distillation.

## 1. Introduction

Convolutional neural networks (CNNs) are widely used in medical image processing due to their strength in feature extraction and classification [1]–[5]. CNNs require a large number of labeled training data to be effective. In the context of medical image processing, access to labeled datasets are limited due to the privacy and regulatory constraints.

Transfer Learning (TL) approaches rely on knowledge obtained from solving one problem, to solve another problem. This means model parameters from a pre-trained model can be transferred to a new model where extensive training data may be lacking [6]–[8]. In TL, the two models must have a similar structure and architecture, which restricts the use of a predefined model.

Knowledge Distillation (KD) was introduced in 2015 as a technique to transfer knowledge of a model to another [9]. KD can be used between models with different structures, addressing a major shortcoming of Transfer Learning. Specifically, knowledge from a complex model (the *teacher*) is transferred to a simpler model (the *student*) by soft labels [10].

Knowledge distillation has interesting applications in expanding the training capabilities of a model. However, more investigation is needed to compare Knowledge Distillation and Transfer Learning. An important question is whether it is possible to use KD as an alternative to the TL. To answer this question, we investigate the application of KD as an alternative for the TL. Our study aims to design a method which has two advantages: (1) an appropriate knowledge transfer technique from a base network to another network (network under transfer), and (2) Designing the model under transfer, with an arbitrary structure. To extract most of the knowledge contained in the teacher model, we also employ unlabeled data during knowledge distillation.

It is possible to create a simple model using the proposed method, which utilizes enough transferred knowledge. To the best of our knowledge, the proposed method is the first study that aims to compensate for the effects of labeled data deficiency in small models used for medical image processing. This method has exciting applications for developing low complexity models that will be implemented in embedded medical imaging devices with low resource budgets. We evaluated our approach using a comprehensive set of experiments for the classification of diabetic retinopathy (DR) images. DR classification is an application area where sufficient training data does not exist, and our approach could offer significant advances in this domain.

The main contributions of this study can be summarized as follows. First, we proposed a novel approach to use knowledge distillation to transfer knowledge of a complex model, which have many learned parameters, into a simple model. Secondly, we use unlabeled data to transfer enough knowledge to a simple model without extra training. Thirdly, we demonstrate our method's effectiveness by designing a simple and efficient network for analysis of Diabetic Retinopathy that can be embedded in medical imaging devices.

The remainder of this paper is structured as follows. Previous studies in DR classification are briefly described in Section 2. In Section 3, the proposed method for knowledge transfer using an unlabeled dataset is explained. In Section 4, and Section 5, experimental results and discussion are presented, respectively. Finally, Section 6 is dedicated to the conclusion of this study.

## 2. Diabetic Retinopathy Classification

Diabetes is a common disease that could harm the micro-vessels in the human eye retina [11-15]. The advanced stage of this disease can lead to diabetic retinopathy (DR), which is considered a prevalent cause of vision loss. Regular retinal monitoring by an expert can be used to prevent vision loss, which is difficult due to its cost and lack of expert accessibility [16].

Automatic screening and analysis of the retina can be considered as a solution to this problem. The processing of retinal images is conducted based on different methods and techniques. Some examples of methods used classification of retinal images include support vector machine (SVM) [17], [18], K-nearest neighbors (KNN) [19], [20], and random forest [21].. Among different methods for automatic screening of the retina, the use of CNNs is probably the best approach. CNNs can employ high-level features to map input images to the output. In [22], DR detection is realized by semantic segmentation of microaneurysms using a CNN. In [23], red lesions are localized by using CNN working on image patches. After applying image processing methods such as image enhancement, DR is analyzed with a CNN structure.

From the perspective of model complexity, different networks are proposed in the literature. For example, in [24]–[26], multiple network structures are utilized, which work either parallel or sequentially. Each network could have a part of the image as its input. In [2], [27], [28], VGG based networks are proposed for DR classification. DR detection requires a structure with a strong feature extracting ability; hence, in [2], [28], VGG network parameters are enhanced using transfer learning from a VGG model pre-trained on Image-Net dataset [29], [30]. Since pre-trained structures are available in the form of a VGG network, [2], [28] were obliged to use a structure such as VGG. By reviewing different CNN structures used for DR analysis, it can be seen that slightly complex networks are employed in many studies. Moreover, we can say that there is no framework for designing a simple structure that can be enriched by the knowledge of complex models.

## 3. Proposed Method

The proposed method is based on three techniques, including transfer learning, knowledge distillation, and employing unlabeled data. In Fig. 1, the structure of the proposed method is illustrated, which contains two main parts, including teacher modeling and student modeling. A teacher model is a temporal model used to train the final student model, and at the inference time, only the student model is active. In the following, the proposed method is explained in detail.

### 3.1. Preprocessing

Datasets used for DR analysis have many images captured in different conditions. Hence, there can be some significant problems that have unfavorable influences on model training. One of the most important ones is the existence of several inappropriate and imperfect images among the dataset. There might be a lack of sufficient contrast for a vast number of samples. Furthermore, images of a dataset may not be balanced between all the existed classes. To address the mentioned problems, one of the solutions can be the elimination of the low-contrast images from the training process. Moreover, the same number of images from different classes can be used for model training to yield better balancing.

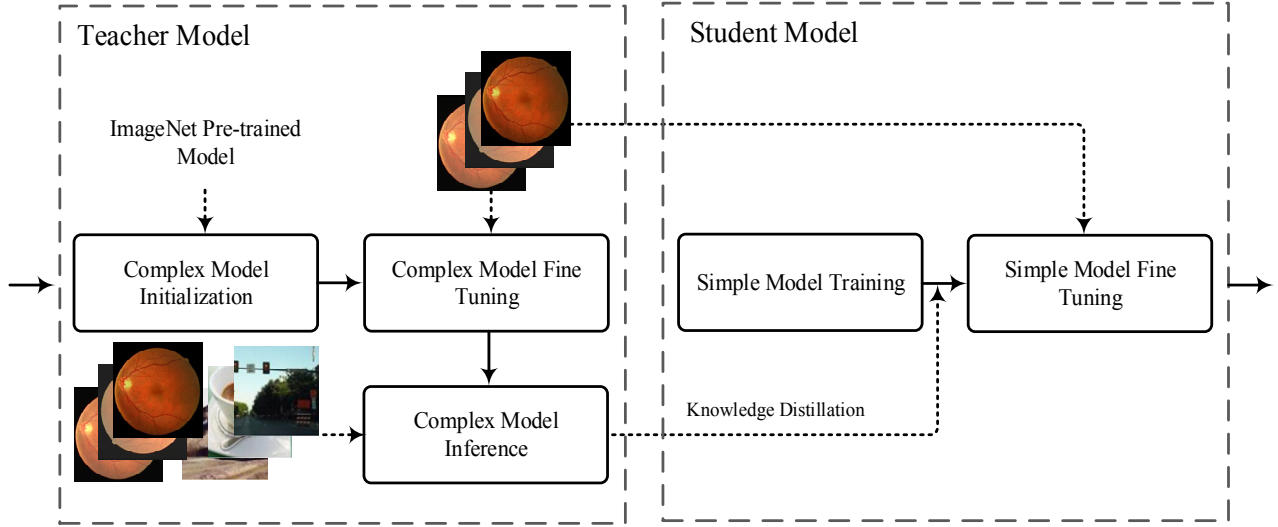

Fig. 1. Block diagram of the proposed method.

The elimination process is conducted for the image's standard deviation, where the images are transformed into the grayscale mode. It has been observed through our experiences which overall standard deviation can be used to demonstrate the contrast of retinal image samples such as the EyePACS dataset [31]. After transforming images into the grayscale mode, the standard deviation is calculated for each image, as illustrated in Equation (1):

$$StdDev_{IMG} = \sqrt{\frac{\sum_{i=1}^{N}(pixel_i - \mu)^2}{N}}$$

$$\mu = \frac{\sum_{i=1}^{N} pixel_i}{N}$$

(1)

In which N is the number of pixels included in image IMG.

Before training the network structure, the application of preprocessing and augmentation can be useful for better training. For preprocessing, the same method as performed in [2] is utilized. Histogram equalization of the retinal images increases the contrast of the vessels, especially micro-vessels, and better represents abnormal regions for DR classification. For preprocessing, local histogram equalization is performed separately on each input channel. For augmentation, the equalized image, row-wise, and column-wise flipping are used to increase our training set.

### 3.2. Teacher model

In the teacher modeling stage, a complex model is trained for the DR classification. For training the teacher model, a VGG structure is considered as the teacher. VGG is selected because its pre-trained version on ImageNet is available. At first, a pre-trained VGG model, trained on ImageNet, is used to initialize the teacher model. After that, the target augmented dataset is fed to the teacher model, and the teacher model is trained on the target dataset. In this way, a network with general feature extraction capability specialized on the target dataset is resulted. In this stage, the network structure is ready for knowledge distillation. Thanks to the distillation process, it is possible to train a model in which its structure is different from the teacher model. We need to determine whether it is possible to transfer all the knowledge of a teacher to a student through the distillation.

A simple and intuitive experiment is performed to answer the above ambiguity. In this experiment, a VGG network as the teacher and another VGG network as a student model is considered. The teacher model is initialized by a VGG model, which is pre-trained on the ImageNet dataset. The teacher is fine-tuned using a retina dataset aiming to classify them for DR levels. The student model is trained using the distilled knowledge from the teacher model. Moreover, for better comparison, a VGG model is trained directly on the same retina dataset for classification of DR levels. The results of these models are illustrated in Fig. 2. It can be observed that the VGG model with distillation has slightly better accuracy than the VGG model without any distillation. Indeed, the student model has a lower accuracy far from the teacher's accuracy.

This observation implies that all of the knowledge of a network may not be transferred through distillation. Explicitly, it can be stated that the knowledge which is transferred to the teacher model is not transferred to the student model through distillation. The accuracy of a simple model can be improved using knowledge distillation [32], [33]. Also, knowledge distillation works better in conditions with limited training data as stated in [34], [35]. With limited training data, knowledge is not transferred even to a model with the same size as the teacher model. During the knowledge distillation, the student only observes the data in the target dataset and is trained based on the corresponding teacher soft labels. The knowledge related to a pre-trained model, and the knowledge that is being used by the transfer-learning can be extracted by observing the images associated with that pre-trained model.

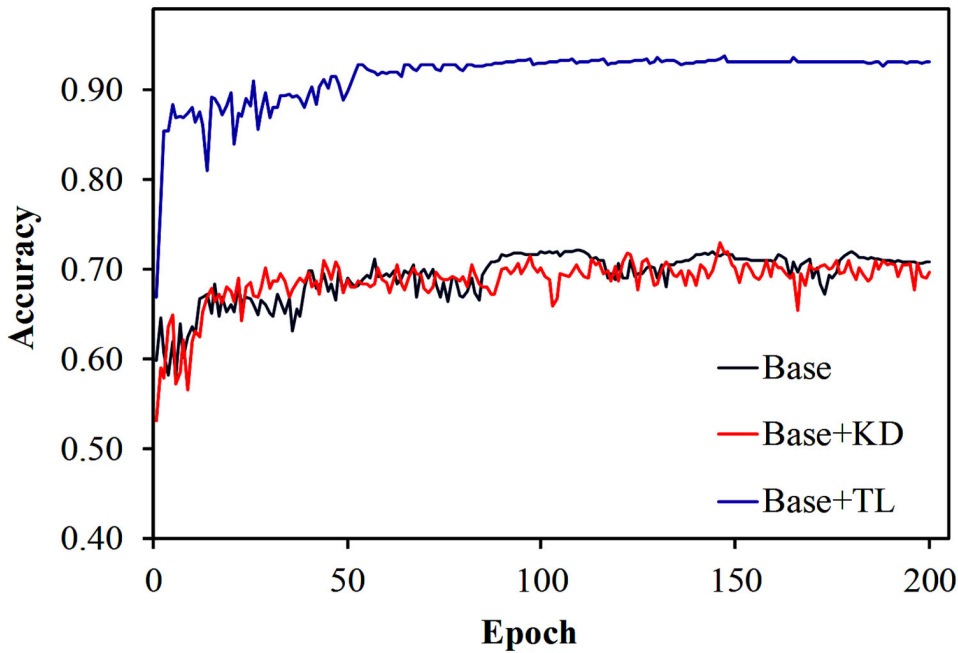

Fig. 2. Comparison of different training VGG models for DR classification. VGG with transfer learning: Base + TL, VGG with knowledge distillation: Base + KD.

Now let us look at the teacher model, which is ready for the transfer-learning. In the proposed method, better knowledge transfer is performed using the labels of the retina images and labels of random natural images. The teacher model set labels on random images that are unlabeled at first. In this way, the student model extracts the knowledge of the teacher model from other images. As illustrated in Fig. 1, after fine-tuning the teacher model, knowledge for distillation is provided from both of the DR and random images

### 3.3. Student model

In the proposed method, we are going to train a simple model as a student such that maximum information from a teacher model could be utilized. Indeed, training a simple model directly by a massive dataset such as ImageNet is a formidable task. Therefore, using unlabeled data could be an alternative to training on a large data set.

In this regard, the student model is selected as a simple model that employs the teacher's knowledge as much as possible. As illustrated in Fig. 1, after training the teacher, the student is trained in two steps. The student follows the same training trend as conducted in the teacher model. In the first step, the student is trained based on the knowledge of the random images using knowledge distillation, which simulates the teacher model's transfer learning. After that, the main images are used to fine-tune the student model using soft labels produced by the teacher model. This stage also simulates the fine-tuning of the teacher model on the main images. In this way, in addition to the training the student on the main images, it is trained based on the other knowledge, which is embedded in the teacher model. Finally, the student can be fine-tuned again using hard labels of the main images.

### 3.4. Proposed methods in the form of pseudo-code

In Fig. 3, the proposed method is represented using pseudo-code. As illustrated in Fig. 3, the procedure is defined clearly in eight consecutive lines. The modules which are utilized in the algorithm are named as Preprocessing and Knowledge Distillation. The preprocessing module takes raw images from the target dataset as input. After selecting well-contrasted images and applying augmentation, including contrast enhancement and brightness improvement, produces the final dataset, which is appropriate for training the networks and return this dataset as an output. The complex model is loaded with a pre-trained model and fine-tuned on the final dataset. The knowledge distillation module takes random unlabeled images and the complex model as inputs. After feeding each unlabeled image to the network, the module assigns the network prediction (known as soft label or logit) to the image's label. Eventually, label-assigned images are generated in the form of a dataset. After making a dataset from the unlabeled images, the simple model is trained on them and finally fine-tuned on the final dataset.

| |
|---|
| **Algorithm1** |
| *Inputs:* |
|     $D_{RT}$: Raw Target Dataset |
|     $S_R$: A Set of Random Unlabeled Images |
|     $S_W$: A Set of Weights correspond to Pre-Trained Model on Big Dataset |
| *Definitions*: |
|     $C_N$: Complex model conforming to $S_W$ |
|     $S_N$: Simple model |
|     $D_R$: Labeled Dataset ← EMPTY |
|     $D_{PT}$: Preprocessed Target Dataset ← EMPTY |
|     $T_C$: Threshold for contrast |
| **Module1: Preprocessing** ($D_{RT}, T_C$) |
|     $D_{PT}$ ← EMPTY |
|     **For** all images, I, in $D_{RT}$ |
|         **If** contrast of $I > T_C$: |
|             apply augmentation on I |
|             add I to $D_{PT}$ |
|         **End If** |
|     **End for** |
|     **Return** $D_{PT}$ |
| **Module2: Knowledge Distillation** (Network, $S_R$) |
|     $D_R$ ← EMPTY |
|     **For** all images, I, in $S_R$ |
|         set Network prediction as the label of I |
|         add I to $D_R$ |
|     **End for** |
|     **Return** $D_R$ |
| 1: **Start** |
| 2: $D_{PT}$ ← Module1($D_{RT}, T_C$) |
| 3: Initialize $C_N$ with $S_W$ |
| 4: Fine-tune $C_N$ using $D_{PT}$ |
| 5: $D_R$ ← Module2($C_N$, $S_R$) |
| 6: Train $S_N$ using $D_R$ |
| 7: Fine-tune $S_N$ using $D_{PT}$ |
| 8: **End** |

Fig. 3. Pseudo-code. of the proposed method.

## 3.5. Knowledge distillation formulation

Method of knowledge distillation has a vital role in the teacher's proper transfer of knowledge to the student. In [36], a teacher-student model with a conditional method is implemented, where teacher's predictions are compared with the original labels. If the prediction is correct, then soft labels are used for distillation; otherwise, hard labels are used for that purpose. In the proposed method, we use conditional distillation. Suppose that we have a teacher $T$ with parameters $w_T$ and a student model $S$ with parameters $w_S$. A set of training sample $D = \{d_1, d_2, ..., d_N\}$, and corresponding labels $L = \{l_1, l_2, ..., l_N\}$ with ($l_i \in \mathbb{R}^{|C|}$) on DR classification as a target dataset is considered and C is the set of all possible classes of $l_i$. Also a set of random images $R = \{r_1, r_2, ..., r_M\}$ without any labels are considered. Two losses can be defined based on Kullback-Leibler (KL) divergence [36]. In KL divergence in cases in which student attempts to approximate teacher's predictions, the teacher's parameters are regarded fixed. Accordingly, the first loss is due to the unlabeled data, which is formulated in the following Equation:

$$L(w_S)_1 = \frac{-1}{M} \sum_{i=1}^{M} \sum_{j=1}^{C} p(r_i:j|T:w_T) \times \log(p(r_i:j|S:w_S)) \qquad (2)$$

In Equation (2), $T:w_T$ and $S:w_S$ represent the teacher network including parameters of $w_T$ and student network including parameters of $w_S$ respectively. The symbol $r_i:j$ stands for consideration of label $j$ for image $r_i$. Accordingly, $p(r_i:j|T:w_T)$ stands for the probability of label $j$ for image $r_i$ predicted by network $T$, and in the same way for $p(r_i:j|S:w_S)$. The second loss also can be defined due to the labeled data, which is conditional as Equation (3):

$$\begin{aligned} L(w_S)_2 = \frac{-1}{N} \sum_{i=1}^{N} \Bigg[ & \Delta\left(\underset{c \in C}{Argmax}(p(d_i:c|T:w_T)) == l_i\right) \\ & \times \left(\sum_{j=1}^{|C|} p(d_i:j|T:w_T) \times \log(p(d_i:j|S:w_S))\right) \\ & + \Delta\left(\underset{c \in C}{Argmax}(p(d_i:c|T:w_T)) \neq l_i\right) \times \log(p(d_i:l_i|S:w_S)) \Bigg] \end{aligned} \qquad (3)$$

The first term of summation indicates the loss due to the samples in which the teacher correctly predicts their labels. The second term indicates the loss of the samples, which are not correctly predicted by the teacher. In Equation (3), $\Delta(x)$ is an indicator function, which is 1 when $x$ is true and 0 when $x$ is false. During training the student model, at first, the student is trained base on the $L(w_S)_1$ to improve the model training capability. At second, the student is fine-tuned using $L(w_S)_2$.

# 4. Experimental Results

Experimental results are conducted in the case of DR classification in retina images. All of the models for DR detection are implemented by Python using the TensorFlow framework. A computer with an Nvidia GPU1080 Ti, and 11GB internal memory is used to implement the proposed method and to train and test different models.

## 4.1. Datasets

Two datasets, including Messidor and EyePACS, are used for our experiments [31], [37]. In the Messidor dataset, there are 1200 RGB images, which we resized them to 300×300. Since the more training samples, the more model generality, after enhancement, by augmentation, we increased the number of images to 4800.

The EyePACS dataset contains about 35,000 images with different sizes. Some images in EyePACS have a dark area around their borders, which could be harmful to model training. The dark area of these images are cropped, and all of them are resized to 300×300. Cropping and resizing can be useful for having a fast training with lower resource consumption. Contrast enhancement and removing images with contrasts lower than a threshold from the training process would yield a better performance model. A vast number of images with visually sufficient contrasts are selected to determine the threshold. The average standard deviation of all images is set as the threshold. In the EyePACS dataset, images in which their standard deviation is less than the threshold are eliminated from the dataset. Hence, 35126 images of dataset decrease to 25231, which 5143 and 20088 images have labels 1 and 0, respectively. Furthermore, in order to balance the number of image labels seen by the model, applying a balance between the numbers of different classes is essential. For the unlabeled data, a set of natural images are randomly selected from the internet containing 20,000 images, which are resized to 300×300. Images of the unlabeled set are fed to the teacher model, as illustrated in Algorithm 1 to set a label to them.

## 4.2. Evaluation Metrics

DR classification accuracy, area under the curve of ROC (receiver operating characteristic curve), and MCC (Matthews Correlation Coefficient) are used to evaluate the performance of different structures. MCC and accuracy are used as equations (4-5), in which TP, TN, FP, and FN represent true positive, true negative, false positive, and false negative, respectively. A Five-fold cross-validation method is used to have a comprehensive validation. For classification, we follow the same definition of DR grading levels, as used in [2], [28].

$$MCC = \frac{(TP \times TN) - (FP \times FN)}{\sqrt{(TP + FP) \times (TP + FN) \times (TN + FP) \times (TN + FN)}} \quad (4)$$

$$Accuracy = \frac{TP + TN}{TP + FP + TN + FN} \quad (5)$$

### 4.3. Models

For the teacher model, a VGG model was employed. This model is not able to yield acceptable results on its own. This problem can be due to the lack of data for the training and weak feature extraction capabilities that by using only DR data occurred. In [2], [28], transfer learning is used to improve their results. By employing the transfer learning technique, it is possible to provide a better learning capability. To this aim, parameters of a VGG network which are pre-trained on ImageNet are used to initialize the teacher parameters.

For the student model, designing small network structures are under consideration. Small structures are different from the VGG network, which means that it is not possible to use a pre-trained VGG model. Moreover, training small models directly on the ImageNet can be a very time-consuming process with a lot of hardware resources. In this experiment, small models are enriched using transfer learning, knowledge distillation, and using unlabeled data.

Two small versions of the VGG network with 16 layers are used, including VGG/4 and VGG/2, in which the number of their filters are divided by 4 and 2, respectively. Also, for better evaluation, a random and small structure with ten convolutional layers is utilized, which is called the "SimpleA" network. The SimpleA network has 20, 20, 30, 30, 40, 40, 160, 160, 250, 250 convolutional filters, in its layers. The conventional training on DR images is named as "Base," learning using knowledge distillation is named as "KD," and employing unlabeled data is called "UL." For better comparison, two structures, including LeNet-like and AlexNet-like, are selected from the [38] and [39], respectively. These models are trained, and corresponding results are reported.

### 4.4. Detection Performance

In Fig. 4 and 5, the results of different training methods for DR classification are illustrated for Messidor and EyePACS datasets, respectively. Fig. 4(a) and Fig. 5(a) are related to the results of the AlexNet-like network, and Fig. 4(b) and Fig. 5(b), are associated with the VGG/4 network. It can be observed that using unlabeled data have an important effect on better training of simple networks.
We can assert that using only knowledge distillation, in different models and datasets, slightly improves the network accuracy. Using unlabeled data leads to a suitable improvement of accuracy in all of the models and datasets. Also, simultaneously using knowledge distillation and unlabeled data, slightly better results are observed in comparison with using only unlabeled data. Fig. 4 and 5 demonstrate that using unlabeled data could improve the training capability of a model.

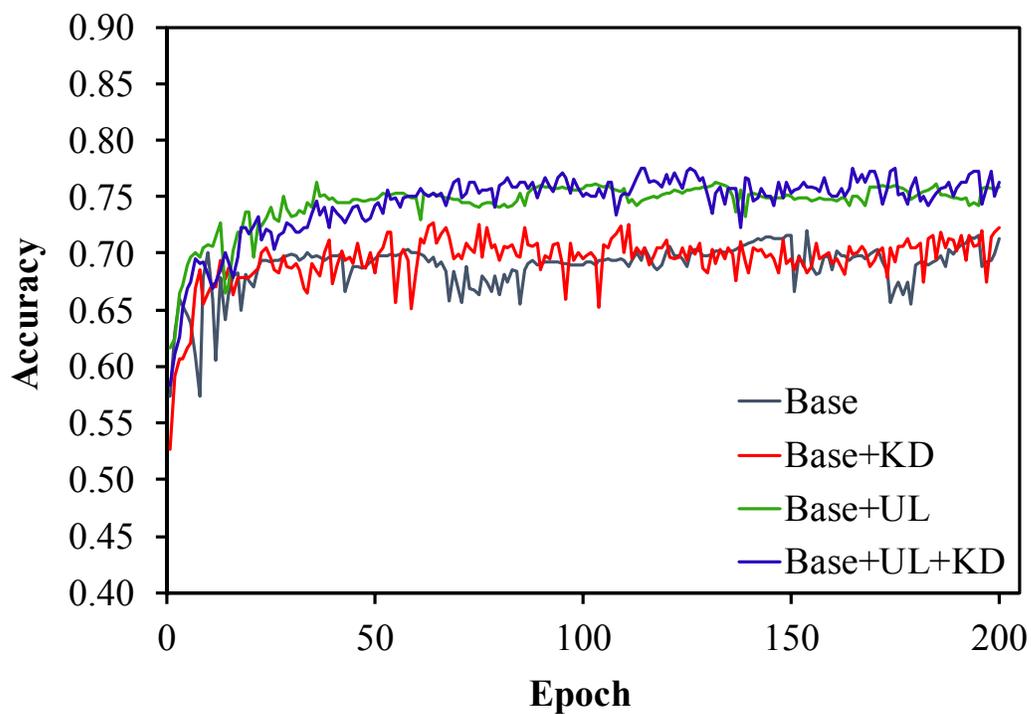

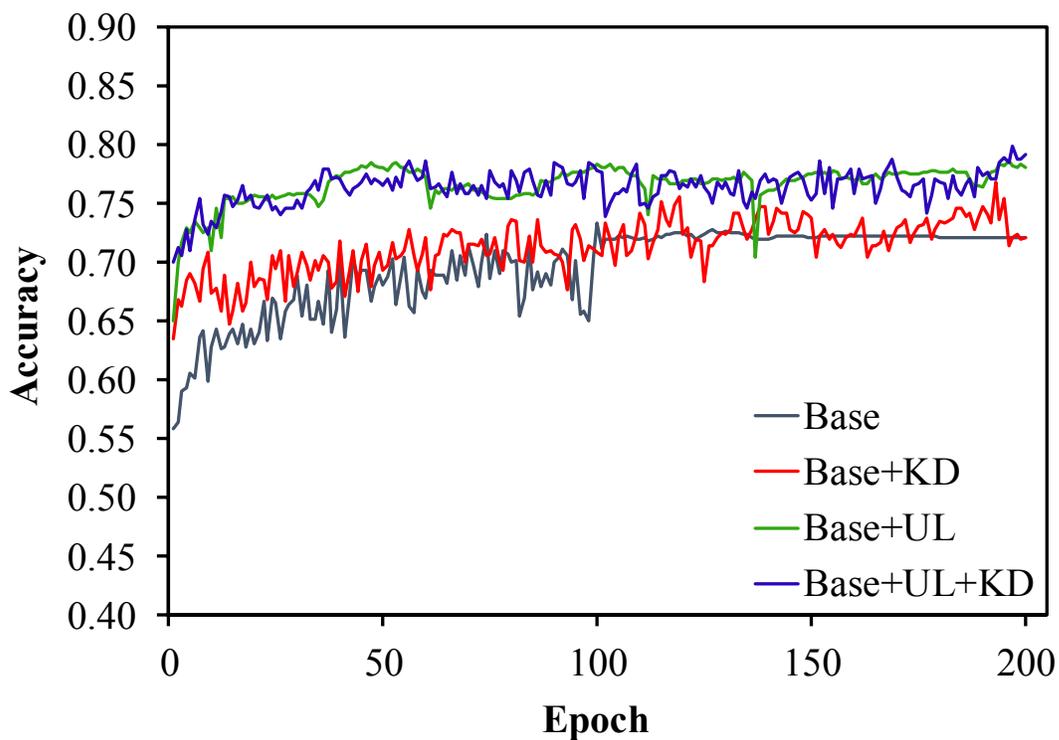

Fig. 4. Accuracies of different training methods for DR classification on Messidor dataset; (a) AlexNet-like, (b) VGG/4

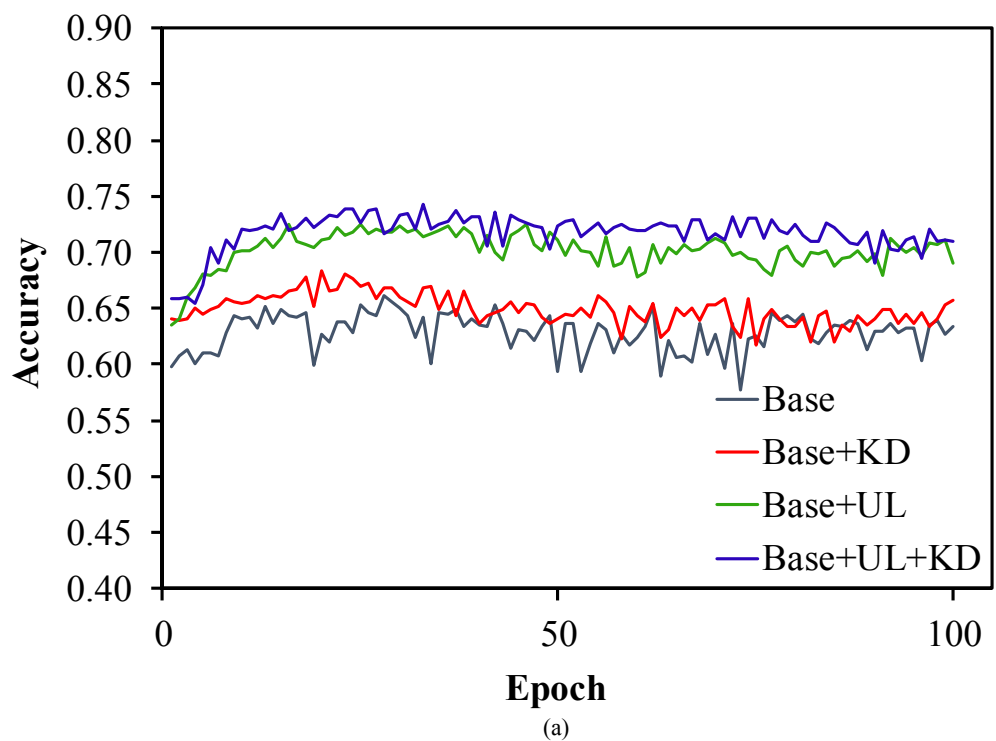

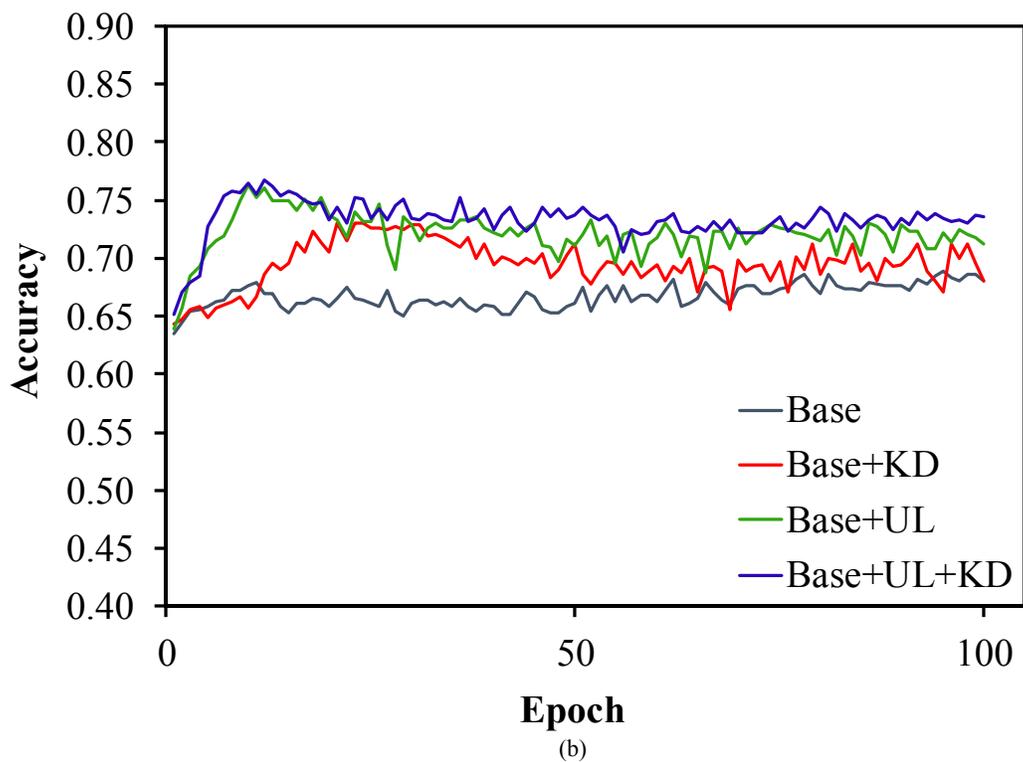

Fig. 5. Accuracies of different training methods for DR classification on EyePACS dataset; (a) AlexNet-like, (b) VGG/4

For a better comparison of different methods, three mentioned networks, as well as those from [38], [39], are trained for 150 epochs, and their detection performances are reported. In Tables 1 and 2, the results of AUC for Messidor and EyePACS dataset are reported respectively, where the best results are bolded. It is observed that for both of Messidor and EyePACS datasets, in all of the models, using unlabeled data causes a significant improvement in the AUC results. In Tables 3, 4, results of detection accuracy for Messidor and EyePACS datasets are reported which similar results are observed. The improvements are also observed in MCC results for both of the employed datasets, as illustrated in Tables 5 and 6.

From Tables 1, 2, 3, 4, 5, 6, it can be concluded that by using unlabeled data detection accuracy and MCC as well as the AUC in all of the simple networks, are improved. Significant differences are observed between the performance of basic training and training using unlabeled data. Finally, we can say that, by knowledge distillation and using unlabeled data, knowledge of a network can be transferred efficiently from a teacher model to a student model.

Table 1. AUCs of different methods for DR detection on Messidor

| Training Methods | Networks | | | | |
|---|---|---|---|---|---|
| | VGG/4 | VGG/2 | SimpleA | LeNet-like [38] | AlexNet-like [39] |
| Base Model | 78.60 | 80.18 | 78.52 | 77.36 | 76.96 |
| Base Model+KD | 83.63 | 81.59 | 79.49 | 78.18 | 78.57 |
| Base Model+UL | 85.17 | 87.09 | 83.34 | 82.10 | **83.07** |
| Base Model+UL+KD | **86.65** | **88.92** | **85.95** | **83.05** | 82.46 |

Table 2. AUCs of different methods for DR detection on EyePACS

| Training Methods | Networks | | | | |
|---|---|---|---|---|---|
| | VGG/4 | VGG/2 | SimpleA | LeNet-like [38] | AlexNet-like [39] |
| Base Model | 69.41 | 72.66 | 71.44 | 65.86 | 66.81 |
| Base Model+KD | 75.25 | 77.95 | 78.75 | 67.48 | 69.25 |
| Base Model+UL | 79.05 | 76.33 | 79.52 | 72.94 | 73.27 |
| Base Model+UL+KD | **79.51** | **78.96** | **80.71** | **73.78** | **76.04** |

Table 3. Accuracies of different methods for DR detection on Messidor

| Training Methods | Networks | | | | |
| --- | --- | --- | --- | --- | --- |
| | VGG/4 | VGG/2 | SimpleA | LeNet-like [38] | AlexNet-like [39] |
| Base Model | 73.37 | 73.58 | 72.11 | 71.84 | 71.95 |
| Base Model+KD | 76.84 | 75.37 | 73.89 | 71.42 | 72.74 |
| Base Model+UL | 78.63 | 79.58 | 76.95 | 76.55 | 76.21 |
| Base Model+UL+KD | **79.89** | **82.32** | **79.16** | **77.05** | **77.58** |

Table 4. Accuracies of different methods for DR detection on EyePACS

| Training Methods | Networks | | | | |
| --- | --- | --- | --- | --- | --- |
| | VGG/4 | VGG/2 | SimpleA | LeNet-like [38] | AlexNet-like [39] |
| Base Model | 68.85 | 71.32 | 70.50 | 66.68 | 67.4 |
| Base Model+KD | 73.04 | 74.48 | 74.84 | 67.56 | 68.36 |
| Base Model+UL | 76.40 | 74.88 | 75.84 | 71.44 | 72.48 |
| Base Model+UL+KD | **76.68** | **76.72** | **76.84** | **72.32** | **74.28** |

Table 5. MCC of different methods for DR detection on Messidor

| Training Methods | Networks | | | | |
| --- | --- | --- | --- | --- | --- |
| | VGG/4 | VGG/2 | SimpleA | LeNet-like [38] | AlexNet-like [39] |
| Base Model | 45.28 | 45.71 | 43.09 | 42.66 | 42.62 |
| Base Model+KD | 52.62 | 49.62 | 46.37 | 42.59 | 44.12 |
| Base Model+UL | 56.20 | 58.14 | 52.79 | 51.98 | 51.12 |
| Base Model+UL+KD | **58.79** | **63.79** | **57.26** | **53.17** | **53.98** |

Table 6. MCC of different methods for DR detection on EyePACS

| Training Methods | Networks | | | | |
|---|---|---|---|---|---|
| | VGG/4 | VGG/2 | SimpleA | LeNet-like [38] | AlexNet-like [39] |
| Base Model | 30.04 | 34.48 | 33.44 | 23.19 | 25.68 |
| Base Model+KD | 39.62 | 43.69 | 43.46 | 25.40 | 28.37 |
| Base Model+UL | 46.93 | 42.84 | 46.15 | 36.44 | 37.60 |
| Base Model+UL+KD | **47.31** | **47.46** | **48.11** | **37.85** | **41.98** |

### 4.5. Complexity Analysis

Several works have investigated DR detection using deep neural network structures. However, they used models with a vast number of parameters requiring many hardware resources. In order to have a better insight into the number of model's parameters and amount of memory requirement, a complexity analysis is conducted. Complexity analysis is done on the VGG model used in [27], [28], AlexNet-like structure in [39], LeNet-like structure in [38], and our three designed simple networks. The results of this analysis for these structures are presented in Fig. 6. Complexity analysis is conducted based on two important perspectives of complexity that existed in the CNN processing. These perspectives are including computational complexity as well as memory complexity. Computational complexity indicates the number of necessary computations for CNN processing, which is related to the number of parameters.

From another perspective, the amount of memory required for storing the intermediate feature maps could be considered as the most important and challenging factor for CNN complexity [40], [41]. The details of computing complexity for one of our designed model (SimpleA) with 300×300 input image is illustrated in Table 7. As illustrated in Table. 7, the SimpleA structure has parameters and feature maps that constitute the computational and memory part of its complexity, respectively.

The main goal of this study is designing models with low complexity structures with a high capability in knowledge transfer. We used simple models and granted them an appropriate knowledge transfer capability. In Fig. 6, the complexity of the models employed in this study is compared with the VGG model as the widely adopted model for DR detection. It can be observed that the model complexities, including computational and memory, are significantly lower in the designed models. Although these models are simple with weak training capabilities, their detection performances are improved using the proposed method.

Table. 7. Structure of SimpleA model and details of computing complexity.

| Layer Type | Conv | Conv | Pooling | Conv | Conv | Pooling |
|---|---|---|---|---|---|---|
| Parameters | 540 | 3600 | - | 5400 | 8100 | - |
| Feature Map | $20 \times 300^2$ | $20 \times 300^2$ | $20 \times 150^2$ | $30 \times 150^2$ | $30 \times 150^2$ | $30 \times 75^2$ |
| Memory | 1800000 | - | 450000 | 675000 | - | 168750 |
| Layer Type | Conv | Conv | Pooling | Conv | Conv | Pooling |
| Parameters | 10800 | 14400 | - | 57600 | 230400 | - |
| Feature Map | $40 \times 75^2$ | $40 \times 75^2$ | $40 \times 38^2$ | $160 \times 38^2$ | $160 \times 38^2$ | $160 \times 19^2$ |
| Memory | 225000 | - | 57760 | 231040 | - | 57760 |
| Layer Type | Conv | Conv | Pooling | Dense | Dense | - |
| Parameters | 360000 | 562500 | - | 12800000 | 1024 | - |
| Feature Map | $250 \times 19^2$ | $250 \times 19^2$ | $250 \times 10^2$ | 512 | 2 | - |
| Memory | 90250 | - | 25000 | 512 | 2 | - |

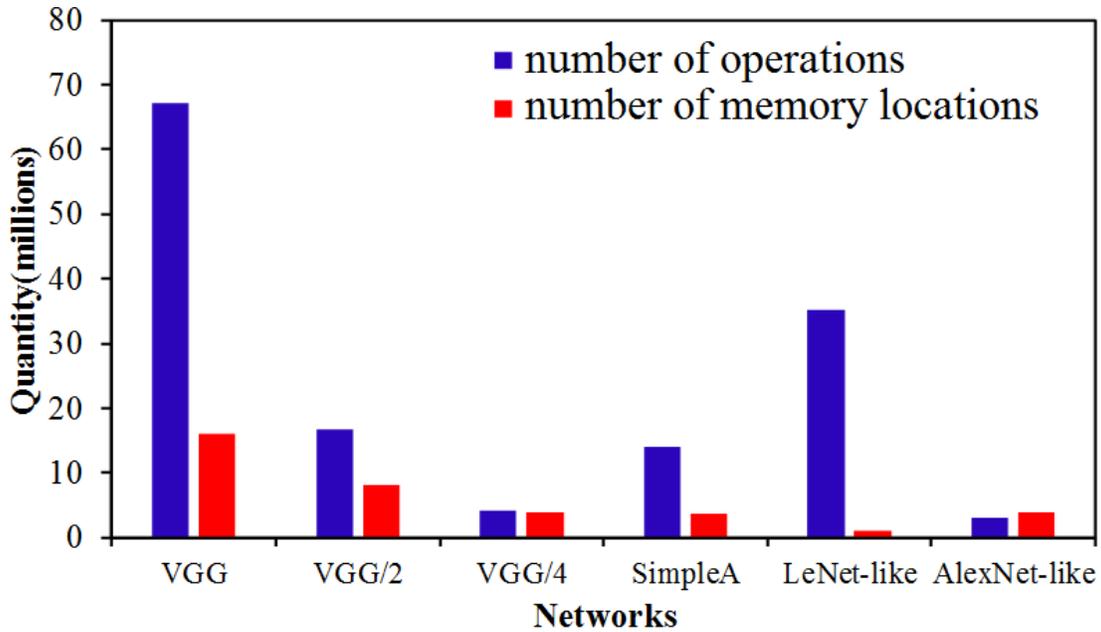

Fig. 6. Complexity of different models

## 5. Conclusion

Considering the outcomes observed in the experimental results section, we see two possibilities that are available by the proposed method. These possibilities include 1) making small trainable models, and 2) easing the training process. Small models have less training capability in comparison with the big models. Sometimes the training weakness of the simple models in an application leads us to not using them. Therefore, using small models can be challenging in applications demanding low complexity models with acceptable performance. It was observed from the results, that although small models were used, an appropriate knowledge from a big and different structure was transferred. The performance of the different small models was improved, and an acceptable DR detection was possible.

From another perspective, we see that this improvement was achieved with a little effort, and with unlabeled data without any expensive annotating procedure. Moreover, unlabeled data are used for training small models to make them better generalize on the new tasks.

Furthermore, training on a big dataset was prevented in the proposed method. For training a DR detection model, it was required a pre-trained model on a large dataset such as ImageNet. Although in the transfer learning, a pre-trained model is used, applying the transfer learning on a new model and structure is not easily possible and has a high cost. Using the proposed method, transfer learning on a new model does not require extra training, and using any pre-trained model is possible. In this way, different models can be used in transfer learning with high knowledge transfer capabilities.

The proposed method could be applied to different applications such as vessel detection in angiograms [42], image compression [43-46], and image fusion [47]. These are applications that may not have large training datasets and the proposed method could be helpful.

In summary, we proposed a new method for knowledge transfer from a complex network to an arbitrary simple network. The proposed algorithm employed the soft labels of a random dataset produced by a complex model to extract all of the model information. This information was used to train a simple model that was not able to perform an appropriate classification. Experimental results, for DR classification, demonstrated that the proposed method for using unlabeled data by simple models improved their accuracy by an average of 6%.

There are applications in medical or general image analysis that portable devices with limited resources have to be used. The proposed approach can be used as a method of knowledge transfer where the available implementation platforms have constraints, the design has to be simple, and the training data is limited.

## REFERENCES

[1]  M. Hajabdollahi, R. Esfandiarpoor, E. Sabeti, N. Karimi, S. M. R. Soroushmehr, and S. Samavi, "Multiple abnormality detection for automatic medical image diagnosis using bifurcated convolutional neural network," *Biomed. Signal Process. Control*, vol. 57, pp. 101792, 2020.

[2]  M. Hajabdollahi, R. Esfandiarpoor, K. Najarian, N. Karimi, S. Samavi, and S. M. Reza

Soroushmehr, "Hierarchical Pruning for Simplification of Convolutional Neural Networks in Diabetic Retinopathy Classification," *IEEE Annual International Conference of the Engineering in Medicine and Biology Society (EMBC)*, pp. 970-973, 2019.

[3]     M. Hajabdollahi, R. Esfandiarpoor, K. Najarian, N. Karimi, S. Samavi, and S. M. Reza-Soroushmeh, "Low complexity convolutional neural network for vessel segmentation in portable retinal diagnostic devices," *IEEE International Conference on Image Processing (ICIP)*, pp.2785-2789, 2018.

[4]     E. Nasr-Esfahani, S. Rafiei, M.H. Jafari, N.r Karimi, J. S. Wrobel, S. Samavi, and S. R Soroushmehr. "Dense pooling layers in fully convolutional network for skin lesion segmentation." *Computerized Medical Imaging and Graphics*, vol(78), p. 101658, 2019.

[5]     Z. Wang and J. Yang, "Diabetic retinopathy detection via deep convolutional networks for discriminative localization and visual explanation," *arXiv preprint* arXiv:1703.10757, 2017.

[6]     M. Oquab, L. Bottou, I. Laptev, and J. Sivic, "Learning and transferring mid-level image representations using convolutional neural networks," *IEEE Computer Society Conference on Computer Vision and Pattern Recognition (CVPR)*, pp.1717-1724, 2014.

[7]     J. Donahue et al., "DeCAF: A deep convolutional activation feature for generic visual recognition," International Conference on Machine Learning (ICML), pp. 647-655, 2014.

[8]     Y. Lecun, Y. Bengio, and G. Hinton, "Deep learning," Nature. vol. 521(7553), pp.436-444, 2015.

[9]     G. Hinton, O. Vinyals, and J. Dean, "Distilling the knowledge in a neural network," arXiv Prepr. arXiv1503.02531, 2015.

[10]    S. Abbasi, M. Hajabdollahi, N. Karimi, and S. Samavi, "Modeling Teacher-Student Techniques in Deep Neural Networks for Knowledge Distillation," arXiv Prepr. arXiv1912.13179, 2019.

[11]    J. Sahlsten et al., "Deep Learning Fundus Image Analysis for Diabetic Retinopathy and Macular Edema Grading," Sci. Rep., vol.9(1), pp.1-11, 2019.

[12]    D. S. W. Ting, G. C. M. Cheung, and T. Y. Wong, "Diabetic retinopathy: global prevalence, major risk factors, screening practices and public health challenges: a review," Clinical and Experimental Ophthalmology, vol.44(4), pp.260-277, 2016.

[13]    M. Hajabdollahi, N. Karimi, S. M. Reza Soroushmehr, S. Samavi, and K. Najarian, "Retinal blood vessel segmentation for macula detachment surgery monitoring instruments," *Int. J. circuit theory Appl.*, vol. 46(6), pp.1166-1180, 2018.

[14]    M. M. Islam, H.-C. Yang, T. N. Poly, W.-S. Jian, and Y.-C. J. Li, "Deep learning algorithms for detection of diabetic retinopathy in retinal fundus photographs: A systematic review and meta-analysis," *Comput. Methods Programs Biomed.*, vol. 191, pp. 105320, 2020.

[15]    P. Chudzik, S. Majumdar, F. Calivá, B. Al-Diri, and A. Hunter, "Microaneurysm detection using fully convolutional neural networks," *Comput. Methods Programs Biomed.*, vol. 158, pp. 185–192, 2018.

[16]    S. Stolte and R. Fang, "A Survey on Medical Image Analysis in Diabetic Retinopathy," *Med. Image Anal.*, vol. 64, pp. 101742, 2020.

[17]    K. M. Adal, P. G. van Etten, J. P. Martinez, K. W. Rouwen, K. A. Vermeer, and L. J. van Vliet,

"An Automated System for the Detection and Classification of Retinal Changes Due to Red Lesions in Longitudinal Fundus Images," *IEEE Transactions on Biomedical Engineering*, vol. 65(6), pp.1382-1390, 2017.

[18] J. Xu *et al.*, "Automatic analysis of microaneurysms turnover to diagnose the progression of diabetic retinopathy," *IEEE Access*, vol. 6, pp. 9632–9642, 2018.

[19] L. Tang, M. Niemeijer, J. M. Reinhardt, M. K. Garvin, and M. D. Abramoff, "Splat feature classification with application to retinal hemorrhage detection in fundus images," *IEEE Trans. Med. Imaging*, vol. 32(2), pp. 364–375, 2012.

[20] M. Niemeijer, M. D. Abramoff, and B. Van Ginneken, "Information fusion for diabetic retinopathy CAD in digital color fundus photographs," *IEEE Trans. Med. Imaging*, vol. 28(5), pp. 775–785, 2009.

[21] U. R. Acharya *et al.*, "Automated diabetic macular edema (DME) grading system using DWT, DCT features and maculopathy index," *Comput. Biol. Med.*, vol. 84, pp. 59–68, 2017.

[22] L. Qiao, Y. Zhu, and H. Zhou, "Diabetic Retinopathy Detection using Prognosis of Microaneurysm and Early Diagnosis System for Non-Proliferative Diabetic Retinopathy Based on Deep Learning Algorithms," *IEEE Access*, vol. 8, pp. 104292 -104302, 2020.

[23] G. T. Zago, R. V. Andreão, B. Dorizzi, and E. O. T. Salles, "Diabetic retinopathy detection using red lesion localization and convolutional neural networks," *Comput. Biol. Med.*, vol. 116, pp. 103537, 2020.

[24] Z. Wang, Y. Yin, J. Shi, W. Fang, H. Li, and X. Wang, "Zoom-in-net: Deep mining lesions for diabetic retinopathy detection," *in Lecture Notes in Computer Science (including subseries Lecture Notes in Artificial Intelligence and Lecture Notes in Bioinformatics)*, pp. 267-275, 2017.

[25] Z. Gao, J. Li, J. Guo, Y. Chen, Z. Yi, and J. Zhong, "Diagnosis of Diabetic Retinopathy Using Deep Neural Networks," *IEEE Access*, vol. 7, pp.3360-3370, 2019.

[26] H. H. Vo and A. Verma, "New deep neural nets for fine-grained diabetic retinopathy recognition on hybrid color space," *IEEE International Symposium on Multimedia (ISM)*, pp. 209-215, 2016.

[27] H. Pratt, F. Coenen, D. M. Broadbent, S. P. Harding, and Y. Zheng, "Convolutional Neural Networks for Diabetic Retinopathy," *in Procedia Computer Science*, pp.200-205, 2016.

[28] Y.-W. Chen, T.-Y. Wu, W.-H. Wong, and C.-Y. Lee, "Diabetic Retinopathy Detection Based on Deep Convolutional Neural Networks," *IEEE International Conference on Acoustics, Speech and Signal Processing (ICASSP)*, pp. 1030–1034, 2018.

[29] A. Krizhevsky, I. Sutskever, and H. Geoffrey E., "Imagenet classification with deep convolutional neural networks," *Adv. Neural Inf. Process. Syst.* pp.1097-1105, 2012.

[30] I. Kandel and M. Castelli, "Transfer Learning with Convolutional Neural Networks for Diabetic Retinopathy Image Classification. A Review," *Appl. Sci.*, vol. 10(6), p. 2021, 2020.

[31] "Kaggle: Diabetic retinopathy detection." [Online]. Available:https://www.kaggle.com/ c/ diabetic -retinopathy-detection, Accessed: 2018-05-14.

[32] A. Mishra and D. Marr, "Apprentice: Using knowledge distillation techniques to improve low-precision network accuracy," in *6th International Conference on Learning Representations, ICLR*

- *Conference Track Proceedings*, 2018.
[33] A. Polino, R. Pascanu, and D. Alistarh, "Model compression via distillation and quantization," *6th International Conference on Learning Representations (ICLR) - Conference Track Proceedings*, 2018.
[34] M. Zhu, K. Han, C. Zhang, J. Lin, and Y. Wang, "Low-resolution Visual Recognition via Deep Feature Distillation," *IEEE International Conference on Acoustics, Speech and Signal Processing - Proceedings (ICASSP)*, pp. 3762-3766, 2019.
[35] G. K. Nayak, K. R. Mopuri, V. Shaj, R. V. Babu, and A. Chakraborty, "Zero-shot knowledge distillation in deep networks," *arXiv Prepr. arXiv1905.08114*, 2019.
[36] Z. Meng, J. Li, Y. Zhao, and Y. Gong, "Conditional Teacher-student Learning," *IEEE International Conference on Acoustics, Speech and Signal Processing - Proceedings (ICASSP)*, pp. 6445-6449. 2019.
[37] E. Decencière *et al.*, "Feedback on a publicly distributed image database: The Messidor database," *Image Anal. Stereol.*, 2014.
[38] M. S. Chowdhury, F. R. Taimy, N. Sikder, and A.-A. Nahid, "Diabetic Retinopathy Classification with a Light Convolutional Neural Network," *International Conference on Comput, Commun, Chem, Mater and Elect Eng (IC4ME2)*, pp. 1–4, 2019.
[39] R. E. Hacisoftaoglu, M. Karakaya, and A. B. Sallam, "Deep Learning Frameworks for Diabetic Retinopathy Detection with Smartphone-based Retinal Imaging Systems," *Pattern Recognit. Lett.*, vol. 135, pp. 409-417, 2020.
[40] V. Sze, S. Member, Y. Chen, S. Member, and T. Yang, "Efficient Processing of Deep Neural Networks : A Tutorial and Survey Efficient Processing of Deep Neural Networks : A Tutorial and Survey," *Proc. IEEE*, vol. 105(12) , Issue: 12 , pp. 2295 - 23292017, 2017.
[41] M. Hajabdollahi, R. Esfandiarpoor, P. Khadivi, S. M. R. Soroushmehr, N. Karimi, and S. Samavi, "Simplification of Neural Networks for Skin Lesion Image Segmentation Using Color Channel Pruning," *Comput. Med. Imaging Graph.*, vol. 82, pp. 101729, 2020.
[42] H. R. Fazlali, N. Karimi, S.M.R. Soroushmehr, S. Sinha, S. Samavi, B. Nallamothu, & K. Najarian, "Vessel region detection in coronary X-ray angiograms," *IEEE International Conference on Image Processing (ICIP)*, pp. 1493-1497, 2015.
[43] N. Karimi, S. Samavi, S.M.R. Soroushmehr, S. Shirani, K. Najarian, K., "Toward practical guideline for design of image compression algorithms for biomedical applications," Expert Systems with Applications, 56, pp. 360-367, 2016.
[44] E. Nasr-Esfahani, S. Samavi, N. Karimi, S. Shirani, "Near lossless image compression by local packing of histogram," IEEE International Conference on Acoustics, Speech and Signal Processing, pp. 1197-1200, 2008.
[45] A. Neekabadi, S. Samavi, S.A. Razavi, N. Karimi, S. Shirani, "Lossless microarray image compression using region-based predictors," IEEE International Conference on Image Processing, vol. 2, pp. II-349, 2007.
[46] M. Nejati, S. Samavi, N. Karimi, S.M.R. Soroushmehr, K. Najarian, "Boosted dictionary learning for image compression," IEEE Transactions on Image Processing, 25(10), pp. 4900-4915, 2016.

[47]   M. Nejati, N. Karimi, S.M.R. Soroushmehr, N. Karimi, S. Samavi, K. Najarian, "Fast exposure fusion using exposedness function," IEEE International Conference on Image Processing (ICIP), pp. 2234-2238, 2017.